\documentclass[12pt,a4paper]{article}

\usepackage{latexsym}
\usepackage{amsmath}
\usepackage{amsfonts}
\usepackage{amssymb}

\newcommand{\be}{\begin{equation}}
\newcommand{\ee}{\end{equation}}
\newcommand{\bea}{\begin{eqnarray}}
\newcommand{\eea}{\end{eqnarray}}

\def\G{{\cal G}}                        %
\def\C{{\cal C}}                       %
\def\cO{{\cal O}}                       %
\def\ri{{\mathrm{i}}}                   %
\def\cL{{\cal L}}                       %
\def\bR{{\mathbb R}}                    %
\def\bC{{\mathbb C}}                    %
\def\bT{{\mathbb T}}                    %
\def\1{{\mbox{\boldmath $1$}}}          %
\def\tr{\mathrm{tr\,}}                  %
\def\diag{\mathrm{diag}}                %
\def\0{{\mbox{\boldmath $0$}}}          %
\def\ext{\mathrm{ext}}                  %
\def\red{\mathrm{red}}                 %
\def\cE{{\cal E}}                       %
\def\cF{{\cal F}}                       %
\def\cD{{\cal D}}                       %
\def\mP{{\cal P}}                       %

\begin{document}

\vspace*{0.5cm}
\begin{center}
{\Large \bf On the duality between the hyperbolic Sutherland  and
the rational Ruijsenaars-Schneider models}
\end{center}

\vspace{0.2cm}

\begin{center}
L. Feh\'er${}^{a,}$
and C. Klim\v c\'\i k${}^b$ \\

\bigskip

${}^a$Department of Theoretical Physics, MTA  KFKI RMKI\\
H-1525 Budapest, P.O.B. 49,  Hungary, and\\
Department of Theoretical Physics, University of Szeged\\
Tisza Lajos krt 84-86, H-6720 Szeged, Hungary\\
e-mail: lfeher@rmki.kfki.hu

\medskip

${}^b$Institut de math\'ematiques de Luminy,
 \\ 163, Avenue de Luminy, \\ 13288 Marseille, France\\
 e-mail: klimcik@iml.univ-mrs.fr

\bigskip

\end{center}

\vspace{0.2cm}

\begin{abstract}

We consider two  families of commuting Hamiltonians on the cotangent
bundle of the group $GL(n,\bC)$, and show that upon an appropriate  {\it single}
 symplectic reduction they descend to  the spectral invariants of the  hyperbolic
Sutherland and of the  rational Ruijsenaars-Schneider  Lax matrices,
respectively. The duality symplectomorphism between these two
integrable models, that was constructed by Ruijsenaars using
direct methods, can be then interpreted geometrically simply as a gauge
transformation connecting two cross sections of the orbits of the reduction group.

\end{abstract}

\newpage

\section{Introduction}
\setcounter{equation}{0}

Around 20 years ago, Ruijsenaars \cite{SR-CMP} undertook the task of constructing action-angle variables
for the non-relativistic and relativistic Calogero models of type $A_n$
(for reviews of these models, see e.g.~\cite{OP-Rep,SR-Kup,Cal,Suth}).
In each case he made use of a commutation relation satisfied by the Lax matrix
of the model under study and another matrix function of the phase-space variables, which he
exhibited directly.
By conjugating these matrices so as to make the Lax matrix diagonal,
he noticed that the conjugate of the other matrix becomes the Lax matrix of another Calogero type
model whose particle-position variables are furnished by the eigenvalues
of the original Lax matrix, i.e., the action variables of the original model.
The so-obtained duality between model-1 and model-2 is thus characterized
by the fact that the action variables of model-1 are the particle-position
variables of model-2, and vice versa.
This observation was used in \cite{SR-CMP}  to derive integration algorithms for the commuting
flows and to calculate  the scattering data.
The simplest manifestation of the duality occurs in the rational Calogero model,
which is actually self-dual \cite{SR-CMP, AKM}.
The self-duality of this model admits a nice  geometric `explanation' in terms
of the symplectic reduction due to
Kazhdan, Kostant and Sternberg \cite{KKS,OP-Rep}.
As it will serve as  a paradigm motivating our considerations,
we next outline this explanation  in some detail.

In fact, Kazhdan, Kostant and Sternberg reduced the cotangent bundle
\be
T^* u(n) \simeq u(n) \times u(n) = \{ (x,y)\}
\label{1.1}\ee
by means of the adjoint action of the group $U(n)$,
 imposing  the moment map constraint
\be
[x,y]= \ri \kappa (\1_n-  w w^\dagger):= \mu_\kappa,
\label{1.2}\ee
where all $n$ components of the column vector $w$ are equal to $1$ and $\kappa$ is a real constant.
The evaluation functions $X(x,y):=\ri x$ and $Y(x,y):=\ri y$
can be viewed as `unreduced Lax matrices' since
$\{ \tr ( X^k)\}$ and $\{\tr ( Y^k)\}$ form two Abelian subalgebras in
the Poisson algebra  $C^\infty(T^*u(n))$.
These Abelian algebras survive the reduction, because their elements
are $U(n)$ invariants.
If one describes the reduced phase space in terms of a gauge slice where $X$ is diagonal,
then -- by solving the moment map constraint -- $Y$ becomes the Lax matrix of the rational
Calogero model whose action variables are
the eigenvalues of $Y$.
If one describes the reduced phase space in terms of a gauge slice where $Y$ is diagonal,
then $X$ becomes the Lax matrix of the `dual Calogero model'.
The correspondence  between the variables of the two Calogero models is obviously
a symplectomorphism, for it represents the transformation between
two gauge slices realizing  the \emph{same} reduced phase space.
The self-duality stems from the symmetrical roles of $x$ and $y$, and
the commutation relation of the Lax matrices is just the constraint (\ref{1.2}) in disguise.

Ruijsenaars hinted in \cite{SR-CMP, RIMS94, RIMS95}  that there
might exist a similar geometric picture behind the duality in other
cases as well, which he left as a problem for `the aficionados of
Lie theory'. Later Gorsky and coworkers \cite{GN,Nekr,Fock+} (see
also \cite{AF,FR}) introduced interesting   new ideas and
 confirmed this expectation in several cases. In particular, they derived the
local version of the  so-called $\mathrm{III}_{\mathrm{b}}$
trigonometric Ruijsenaars-Schneider  model \cite{SR-Kup,RIMS95,DV},
by reducing a Hamiltonian system on the magnetic cotangent bundle
of the loop group of
 $U(n)$ \cite{GN}\footnote{It is worth noting that
 besides the $\mathrm{III}_{\mathrm{b}}$ model there exist also other
 important, physically different real forms \cite{SR-Kup,RIMS95}
 of the complex trigonometric Ruijsenaars-Schneider model.}. The investigations in
\cite{GN,Nekr,Fock+,AF,FR} focused on the local aspects and did not
touch on the quite tricky global definition of the pertinent gauge
slices,
 which is necessary to obtain complete commuting flows.
 We believe, however, that the reduction approach  works also {\it globally}
  and   that it is possible to characterize the Ruijsenaars duality  in  a {\it finite dimensional}
symplectic reduction picture in {\it all}  cases
studied in \cite{SR-CMP,RIMS94,RIMS95}.
We plan to explore this issue systematically in a series of papers,
and here we report the first results of our analysis.

In this paper we  study a case   of the duality which
has not   been previously described    in the symplectic reduction
framework. Namely,
we expound the geometric picture that links together
the dual pair consisting of the hyperbolic Sutherland model
and the
rational Ruijsenaars-Schneider model \cite{SR-CMP}.
In Sect.~2, we
start with two sets of `canonical integrable systems' on the
cotangent bundle of the real Lie group $GL(n,\bC)$, whose
Hamiltonians span two commutative families,  $\{H_j\}$ and
$\{\hat H_k\}$. By `canonical
integrability' we simply mean that one can directly write down the
Hamiltonian flows.  Then, in Sect.~3, we describe a  symplectic
reduction of $T^*GL(n,\bC)$ for which  our canonical
integrable systems  descend to the reduced phase space.
By using the shifting trick of symplectic reduction,
we exhibit two distinguished  cross sections
of the orbits of the gauge group that define two models of the reduced
phase space. In terms of  cross section $S_1$, the family $\{ H_j\}$
translates into the action variables of the Sutherland model and the
family $\{\hat H_k\}$ becomes equivalent to the Sutherland particle
coordinates. Cross section $S_1$ is described by Theorem 1, which
summarizes well-known results \cite{OP-Rep}. In terms of cross
section  $S_2$, the family $\{H_j\}$ translates into the coordinate
variables of the rational Ruijsenaars-Schneider  model and the family
$\{\hat H_k\}$ gives the action variables of this model. We call
cross section $S_2$ the `Ruijsenaars gauge slice'. Its
characterization by Theorem 2 below is our principal
technical result.

The duality symplectomorphism between the hyperbolic Sutherland and the
rational Ruijsenaars-Schneider models
will be realized as the gauge transformation between the cross sections $S_1$ and $S_2$ mentioned above.
In addition, analogously to the case of the rational Calogero model,
the symplectic reduction immediately yields
integration algorithms for the commuting flows of the dual pair of models, and permits to recognize
the commutation relations of the Lax matrices used by Ruijsenaars as equivalents
to the moment map constraint of the reduction.
These consequences of Theorem 1 and Theorem 2 are developed in Sect.~4.

The self-contained presentation of the relatively simple example of the Ruijsenaars duality that follows
may also
facilitate the geometric understanding of this remarkable phenomenon in more complicated cases.

\section{Canonical integrable systems on $T^* GL(n,\bC)$}
\setcounter{equation}{0}

Here we describe the two families of canonical integrable systems and
their symmetries  that will be used to derive  the dual pair of integrable
 many-body models by symplectic reduction.
For a general reference on  symplectic reduction, we mention
 the textbook  \cite{LM}.

Consider the real Lie algebra $\G:= gl(n,\bC)$ and equip it
with the invariant bilinear form
\be
\langle X, Y\rangle := \Re \tr(XY)
\qquad
\forall X,Y\in \G,
\label{2.1}\ee
which permits to identify $\G$ with $\G^*$ by the map $\jmath: \G^* \to \G$ as
\be
\langle \jmath(\alpha), X\rangle = \alpha(X)
\qquad
\forall \alpha\in \G^*,\, X\in \G.
\label{2.2}\ee
Then use left-trivialization to obtain a model of the cotangent bundle of the real Lie group
$G:= GL(n,\bC)$
as
\be
T^* G \simeq G \times \G = \{ (g, J^R)\,\vert\, g\in G,\,\, J^R \in \G\},
\label{2.3}\ee
where $\alpha_g \in T_g^* G$ is represented by $(g, \jmath\circ L_g^*(\alpha_g))\in G\times \G$
with  the left-translation
$L_g\in \mathrm{Diff}(G)$.
In terms of this model, the canonical symplectic form $\Omega$ of $T^*G$ takes the form
\be
\Omega = d \langle J^R, g^{-1} dg\rangle.
\label{2.4}\ee
Next introduce the matrix functions $\cL_1$ and $\cL_2$ on $T^*G$ by the definitions
\be
\cL_1(g,J^R):= J^R
\quad\hbox{and}\quad
\cL_2(g,J^R):= g g^\dagger.
\label{2.5}\ee
We may think of these as `unreduced Lax matrices' since they generate the
Hamiltonians
\be
H_j := \frac{1}{j} \Re\tr(\cL_1^j), \qquad j=1,\ldots, n,
\label{2.6}\ee
and
\be
\hat H_k:= \frac{1}{2k} \tr(\cL_2^k),
\qquad
k=\pm 1,\ldots, \pm n,
\label{2.7}\ee
so that both $\{ H_j\}$ and $\{ \hat H_k\}$ form
commuting sets\footnote{We could have admitted also the imaginary part of the trace in (\ref{2.6}), and the
Hamiltonians in (\ref{2.7}) are not functionally independent; but these are the definitions that
will prove convenient for us.}
with respect to
the Poisson bracket on the phase space $T^*G$.

It is easy to determine the flows of the Hamiltonians introduced above, through any initial
value $(g(0), J^R(0))$.
In fact, the flow belonging to $H_j$ (\ref{2.6})  is given by
\be
g(t) = g(0) \exp(t (J^R(0))^{j-1}),
\qquad
J^R(t)=J^R(0).
\label{2.8}\ee
The flow generated by $\hat H_k$ (\ref{2.7}) reads as
\be
J^R(t)= J^R(0) - t \left(g^\dagger(0) g(0)\right)^k,
\quad
g(t)=g(0).
\label{2.9}\ee

We are going to reduce the phase space $T^*G$ by using the symmetry group
\be
K:=U(n)^L \times U(n)^R,
\label{2.11}\ee
where the notation reflects the fact that the two $U(n)$ factors operate by
left- and right-multiplications, respectively.
This means that an element $(\eta_L, \eta_R) \in K$ (with $\eta_{L,R}\in U(n)$) acts
by the symplectomorphism $\Psi_{\eta_L,\eta_R}$ defined by
\be
\Psi_{\eta_L,\eta_R}(g,J^R) := (\eta_L g \eta_R^{-1}, \eta_R J^R \eta_R^{-1}).
\label{2.12}\ee
This action is generated by an equivariant moment map.
To describe this map, let us note that every $X\in \G$ can be uniquely decomposed (the Cartan
decomposition) as
\be
X= X_+ + X_-
\quad\hbox{with}\quad
X_+ \in u(n),\, X_- \in \ri u(n),
\label{2.13}\ee
i.e., into anti-Hermitian and Hermitian parts.
Identify $u(n)$ with $u(n)^*$ by the `scalar product'
$\langle . , . \rangle$ restricted to $u(n) \subset \G$.
Then the moment map $\Phi: T^* G \to u(n)^L \oplus u(n)^R$ reads
\be
\Phi(g, J^R)= ( (gJ^R g^{-1})_+, - J^R_+).
\label{2.14}\ee

The Hamiltonians $H_j$ and $\hat H_k$ are invariant under the symmetry group $K$.
Hence the commutative character of the families $\{ H_j\}$ and $\{ \hat H_k\}$ survives any
symplectic reduction based on this symmetry group.
It is also clear that the flows of the reduced Hamiltonians will be provided
as projections of the above given obvious flows to the reduced phase space.
However, in general it is a matter of  `art and good luck' to find a value of the
moment map that leads to interesting reduced systems.
In the present case it is well-known \cite{OP-Rep} that by setting the $u(n)^R$-component of the moment map $\Phi$
to zero, and by setting the $u(n)^L$-component equal to the constant $\mu_\kappa$ defined in
(1.2), one obtains the hyperbolic Sutherland model from the Hamiltonian system
$(T^*G, \Omega, H_2)$.

Our goal is to characterize the reduced Hamiltonian systems coming from
$(T^*G, \Omega, H_j)$ and from $(T^*G, \Omega, \hat H_k)$.
To describe  the latter systems, it will be technically very
convenient to make use of the so-called shifting trick of symplectic reduction.
This means that before performing the reduction we extend the phase space
by a coadjoint orbit.
In the present case, we consider the $U(n)$ orbit through $-\mu_\kappa$ given by
\be
\cO^L_{\kappa}:= \{ \ri\kappa (v v^\dagger - \1_n)   \,\vert\, v\in \bC^n,\, \vert v \vert^2 = n\,\}.
\label{2.15}\ee
The orbit carries its own (Kirillov-Kostant-Souriau) symplectic form, which we
denote by $\Omega^\cO$.
The vector $v$ matters only up to phase and  $(\cO^L_\kappa, \Omega^\cO)$ can be identified as a
copy of $\bC P_{n-1}$ endowed with a multiple of the K\"ahler form defined by the
Fubini-Study metric.

\section{Symplectic reduction of the extended phase space}
\setcounter{equation}{0}

The extended phase space to consider now is
\be
T^*G \times \cO_\kappa^L = \{ (g,J^R, \xi) \}
\label{3.1}\ee
with the symplectic form
\be
\Omega^\ext = \Omega + \Omega^\cO.
\label{3.2}\ee
The symmetry group $K$ acts by the symplectomorphisms $\Psi_{\eta_L,\eta_R}^\ext$ given by
\be
\Psi^\ext_{\eta_L,\eta_R}(g,J^R,\xi) := (\eta_L g \eta_R^{-1}, \eta_R J^R \eta_R^{-1}, \eta_L \xi \eta_L^{-1}),
\label{3.3}\ee
and the corresponding moment map $\Phi^\ext$ is
\be
\Phi^\ext(g, J^R,\xi)= ( (gJ^R g^{-1})_+ +\xi , - J^R_+).
\label{3.4}\ee
We are going to reduce at the zero value of the extended moment map,
i.e., we wish to describe the reduced phase space
\be
T^*G \times \cO_\kappa^L//_0 K.
\label{3.5}\ee
In our case this space of $K$-orbits is a smooth manifold, as will be seen from its models.
It is equipped with the reduced symplectic form, $\Omega^\red$, which is characterized by the equality
\be
\pi^* \Omega^\red = \iota^* \Omega^\ext.
\label{3.5+}\ee
Here $\pi$ is the submersion from $(\Phi^\ext)^{-1}(0)$ to the space of orbits in (\ref{3.5}),
and $\iota$ is the injection from $(\Phi^\ext)^{-1}(0)$ into $T^*G\times \cO^L_\kappa$.
Before turning to the details, a few remarks are in order.

First, note that one can reduce in steps, initially implementing
only the reduction by the factor $U(n)^R$ of $K$ (\ref{2.11}). This
leads to the equality \be T^*G \times \cO_\kappa^L//_0 K = [T^*
(G/U(n)^R) \times \cO^L_\kappa ]//_0 U(n)^L, \label{3.6}\ee where
$G/U(n)^R$ can be viewed as the symmetric space of positive definite
Hermitian matrices. It is for computational convenience that we start from
the larger phase space $T^* GL(n,\bC)$.

Second, the advantage of the shifting trick is that convenient models of the reduced
phase space  may become available as cross sections of the $K$-orbits in
$(\Phi^\ext)^{-1}(0)$, which are more difficult to realize without using
the auxiliary orbital degrees of freedom. But in principle one can always do without the shifting
trick: in our case we have $T^*G \times \cO_\kappa^L//_0 K \equiv T^*G //_{(\mu_\kappa,0)} K$,
with the reduced phase space defined by (Marsden-Weinstein)  point reduction on the right-hand-side of the equality.

Third, if we define
\be
\xi(v):= \ri \kappa ( v v^\dagger - \1_n)
\qquad
\forall v\in \bC^n \quad\hbox{with}\quad \vert v \vert^2 =n,
\label{3.7}\ee
then we have
\be
\eta \xi(v) \eta^{-1} = \xi(\eta v)
\qquad
\forall \eta \in U(n).
\label{3.8}\ee
This means that the obvious action of $U(n)$ on $\bC^n$ underlies
the coadjoint action on the orbit $\cO^L_\kappa$, and we shall
use this to transform
the $\bC^n$-vectors $v$ that correspond to the elements $\xi(v)$.

Fourth, we define the extended Hamiltonians $H_j^\ext$ and $\hat H_k^\ext$ by
\be
H_j^\ext(g,J^R,\xi):= H_j(g,J^R),
\qquad
\hat H^\ext_k(g,J^R,\xi):= \hat H_k(g,J^R).
\label{3.9}\ee
The flows on $T^*G \times \cO^L_\kappa$ are obtained from
the flows of the unextended Hamiltonians on $T^*G$ simply by adding the relation $\xi(t)=\xi(0)$.
We may also define the matrix functions
$\cL_1^\ext$ and $\cL_2^\ext$ on $T^*G \times \cO^L_\kappa$ by
\be
\cL_1^\ext(g,J^R,\xi) =  J^R\quad\hbox{and}\quad
\cL_2^\ext(g, J^R,\xi) = g g^\dagger,
\label{3.10+}\ee
 whereby we can write
\be
H_j^\ext = \frac{1}{j} \Re\tr((\cL_1^\ext)^j)
\quad\hbox{and}\quad
\hat H_k^\ext= \frac{1}{2k} \tr((\cL_2^\ext)^k).
\label{3.11+}\ee

Now we turn to the description of two alternative models of the reduced phase space
that will be shown  to carry a dual pair of integrable many-body models.
As a preparation,
 we associate  to any vector $q \in \bR^n$ the diagonal matrix
\be
\mathbf{q}:= \diag(q^1,\ldots, q^n).
\label{3.10}\ee
We let $\C$ denote the open domain (Weyl chamber)
\be
\C:= \{\, q\in \bR^n \,\vert\, q^1 > q^2 >\cdots > q^n\,\},
\label{3.11}\ee
and equip
\be
T^* \C \simeq \C \times \bR^n
\label{3.12}\ee
with the Darboux form
\be
\Omega_{T^*\C}(q,p) := \sum_k dp_k \wedge dq^k.
\label{3.13}\ee

\subsection{The Sutherland gauge slice  $S_1$}

Let us define the $\ri u(n)$-valued matrix function $L_1$ on $T^*\C$ (cf.~(\ref{3.12})) by the formula
\be
L_1(q,p)_{jk}:= p_j \delta_{jk} - \ri (1-\delta_{jk}) \frac{\kappa}{ \sinh(q^j - q^k)}.
\label{3.14}\ee
This is just the standard Lax matrix of the hyperbolic Sutherland model \cite{Mos,CRM,OP-Rep}.
The following result is well-known \cite{KKS,OP-Rep}, but for readability
we nevertheless present it together with a proof.

\medskip
\noindent
\textbf{Theorem 1.}
\emph{
The manifold $S_1$ defined by
\be
S_1:= \{\, (e^{\mathbf{q}}, L_1(q,p), -\mu_\kappa)\,\vert\, (q,p)\in \C\times \bR^n\, \}
\label{3.15}\ee
 is a global cross section of the $K$-orbits
in the submanifold $(\Phi^\ext)^{-1}(0)$ of $T^* G \times \cO^L_\kappa$.
If
 $\iota_1: S_1 \to T^*G \times \cO_\kappa^L$ is the  obvious injection,
then in terms of the coordinates $q$, $p$ on $S_1$ one has
\be
\iota_1^*(\Omega^\ext) = \sum_k dp_k \wedge dq^k.
\label{3.16}\ee
Therefore, the symplectic manifold $(S_1, \sum_k dp_k\wedge dq^k)\simeq (T^*\C, \Omega_{T^*\C})$
is a model of the reduced phase space defined by (\ref{3.5}).}

\medskip
\noindent {\bf Proof.} Our task is to bring every element of
$(\Phi^\ext)^{-1}(0)$ to a unique normal form by the `gauge
transformations' provided by the group $K$. For this purpose, let us
introduce the submanifold $\mP$ of $GL(n,\bC)$ given by the positive
definite Hermitian matrices. Recall that the exponential map from
$\ri u(n)$ to $\mP$ is a diffeomorphism. By the polar (Cartan)
decomposition, every $g\in GL(n,\bC)$ can be uniquely written as
\be
g= g_- g_+ \quad\hbox{with}\quad g_-\in \mP,\, g_+ \in U(n).
\label{3.17}\ee
It is readily seen that $g$ can be transformed by
the $K$-action into an element for which
\be g_+ = \1_n
\quad\hbox{and}\quad g_-= e^{\mathbf{q}} \quad \hbox{with}\quad q\in
\bR^n,\quad q^1 \geq q^2 \geq \cdots \geq q^n.
\label{3.18}\ee
The
moment map constraint requires $J_+^R=0$ and, by (\ref{3.4}), for
triples of the form $(e^{\mathbf{q}}, J_-^R, \xi(v))$ we are left
with the condition
 \be \left(e^{\mathbf q} J_-^R
e^{-\mathbf{q}}\right)_+ + \xi(v)=0.
\label{3.19}\ee
This implies
that the diagonal entries of the Hermitian matrix $J_-^R$ are
arbitrary and the diagonal entries of the anti-Hermitian matrix
$\xi(v)$ vanish. By (\ref{3.7}), it follows from $\xi(v)_{jj}=0$
that
\be v_j = e^{\ri \theta_j}, \qquad \forall j, \label{3.20}\ee
with some phase factors $\theta_j$. The off-diagonal components of
the constraint (\ref{3.19}) are
\be (J_-^R)_{jk}  \sinh (q^j - q^k)
+ \ri \kappa v_j \bar v_k =0, \qquad \forall j\neq k.
\label{3.21}\ee
Since $v_j \bar v_k\neq 0$, we see from (\ref{3.18})
and (\ref{3.21}) that $q$ must belong to the open Weyl chamber $\C$
(\ref{3.11}). Then the residual gauge transformations permitted by
the partial gauge fixing condition (\ref{3.18}) are given precisely
by the maximal torus of $\bT_n \subset U(n)$,  diagonally embedded
into $K=U(n)^L \times U(n)^R$.
They operate according to
\be
(e^{\mathbf{q}}, J^R_-, v) \mapsto (e^{\mathbf{q}}, \tau J^R_- \tau^{-1}, \tau v)
\qquad  \forall \tau\in \bT_n.
\label{3.22}\ee
Hence we can completely fix the residual gauge freedom by
transforming the vector $v$ into the representative $w$, whose components are all
equal to $1$.
At the same time, by (\ref{3.21}), $J_-^R$ becomes equal to $L_1(q,p)$, where $\mathbf{p}$
is the arbitrary diagonal part of $J_-^R$.
The calculation of $\iota_1^* (\Omega^\ext)$ as well as the rest of the statements of the theorem
is now obvious.
\emph{Q.E.D.}

\medskip
\noindent
\textbf{Remark 1.}
If we spell out  the moment map constraint (\ref{3.19}) for the solution
$J_-^R= L_1$, $v= w$ and also multiply this equation
both from the left and from the right by $e^{\mathbf{q}}$, then we obtain
\be
[ e^{2\mathbf{q}}, L_1(q,p)] +
 2 i\kappa\left( (e^{\mathbf{q}} w) (e^{\mathbf{q}} w)^\dagger - e^{2\mathbf{q}}\right)=0.
 \ee
This is the commutation relation for the Lax matrix (\ref{3.14}) used in \cite{SR-CMP}.

\subsection{The Ruijsenaars gauge slice $S_2$}

In this subsection we denote the elements of
$T^*\C=\C\times \bR^n$ as pairs $(\hat p, \hat q)$.
We introduce the $\mP$-valued matrix function $L_2$ on $T^*\C$ by the formula
\be
L_2(\hat p,\hat q)_{jk}=u_j(\hat p,\hat q)
 \left[ \frac{2\ri \kappa}{2 \ri \kappa + (\hat p^j-\hat p^k)} \right] u_k(\hat p,\hat q)
\label{3.23}\ee
with the $\bR_+$-valued functions
\be
u_j(\hat p, \hat q):=  e^{-\hat q_j} \prod_{m\neq j}
 \left[ 1 + \frac{4 \kappa^2}{(\hat p^j - \hat p^m)^2}\right]^\frac{1}{4}.
 \label{3.24}\ee
One can calculate the principal minors of $L_2$ with the help of the Cauchy determinant formula,
and thereby confirm that $L_2$ is  indeed a positive definite matrix.
Therefore it admits a unique positive definite square root, and we use it to define the
$\bR^n$-valued function
\be
v(\hat p, \hat q):= L_2(\hat p, \hat q)^{-\frac{1}{2}} u(\hat p,\hat q),
\label{3.25}\ee
where $u=(u_1,\ldots, u_n)^T$.
It can be verified directly,  and can be seen also from the proof below, that
$\vert v(\hat p, \hat q)\vert^2 =n$.
Thus, by using (\ref{3.7}), we have
the $\cO_\kappa^L$-valued function
\be
\xi(\hat p, \hat q):= \xi(v(\hat p, \hat q)).
\label{3.26}\ee
The function $L_2$ (\ref{3.23}) in nothing but the standard Lax matrix of the rational
 Ruijsenaars-Schneider model \cite{RS}, with variables denoted by somewhat unusual letters.
Using the above notations, we now state the main technical result of the present paper.

\medskip
\noindent
\textbf{Theorem 2.}
\emph{
The manifold $S_2$ defined by
\be
S_2:= \{\, ( L_2(\hat p, \hat q)^\frac{1}{2},\mathbf{\hat p}, \xi(\hat p,\hat q))\,\vert\,
(\hat p,\hat q)\in \C\times \bR^n\, \}
\label{3.27}\ee
 is a global cross section of the $K$-orbits
in the submanifold $(\Phi^\ext)^{-1}(0)$ of $T^* G \times \cO^L_\kappa$.
If
 $\iota_2: S_2 \to T^*G \times \cO_\kappa^L$ is the  obvious injection,
then in terms of the coordinates $\hat p$, $\hat q$ on $S_2$ one has
\be
\iota_2^*(\Omega^\ext) = \sum_k d \hat q_k \wedge d\hat p^k.
\label{3.28}\ee
Therefore, the symplectic manifold $(S_2, \sum_k d\hat q_k\wedge d\hat p^k)\simeq (T^*\C, \Omega_{T^*\C})$
is a model of the reduced phase space defined by (\ref{3.5}).}

\medskip
\noindent
{\bf Proof.}
Let us begin by noting that by means of the $K$-action we can transform each element
of $(\Phi^\ext)^{-1}(0)$ into an element $(g, J^R, \xi(v))$ that satisfies
\be
g=g_- \in \mP
\quad\hbox{and}\quad
J^R = \mathbf{\hat p}
\quad \hbox{with}\quad
\hat p \in \bR^n,\quad
\hat p^1 \geq \hat p^2 \geq \cdots \geq \hat p^n.
\label{3.29}\ee
After this partial gauge fixing the moment map constraint becomes
\be
g_-^{-1} \mathbf{\hat p} g_- - g_- \mathbf{\hat p} g_-^{-1} = 2 \xi(v).
\label{3.30}\ee
In order to solve this equation, we multiply it both from the left and from the right by $g_-$,
which gives
\be
[\mathbf{\hat p} , g_-^2] =  2\ri \kappa ( u u^\dagger - g_-^2),
\label{3.31}\ee
where we have combined the unknowns $g_-$ and $v$ to define
\be
u:= g_- v.
\label{3.32}\ee
This equation then permits us to express $g_-^2$ in terms of $\mathbf{\hat p}$ and $u$ as
\be
(g_-^2)_{jk} = u_j \left[ \frac{2\ri \kappa}{2 \ri \kappa + (\hat p^j-\hat p^k)} \right] \bar u_k.
\label{3.33}\ee
By calculating the determinant from the last relation, we obtain
\be
\det(g_-^2) = \left(\prod_m \vert u_m\vert^2\right)
\prod_{j<k} \frac{(\hat p^j - \hat p^k)^2}{(\hat p^j - \hat p^k)^2 + 4 \kappa^2}.
\label{3.34}\ee
Since $\det(g_-^2) \neq 0$,  we must have
\be
\hat p^1 > \hat p^2 >\cdots > \hat p^n
\quad
\hbox{and}\quad
u_j \neq 0 \quad \forall j.
\label{3.35}\ee
In particular,  $\hat p$ must belong to the open Weyl chamber $\C$.
This implies that the residual gauge transformations permitted by our partial gauge
fixing are generated  by the maximal torus $\bT_n$ of $U(n)$,
diagonally embedded into $K$, which act according to
\be
(g_-, \mathbf{\hat p}, v) \mapsto (\tau g_- \tau^{-1}, \mathbf{\hat p}, \tau v)
\qquad  \forall \tau\in \bT_n .
\label{3.36}\ee
By applying these gauge transformations  to $u$ in (\ref{3.32}) we have
\be
\tau: u \mapsto \tau u.
\label{3.37}\ee
It follows that we obtain a complete gauge fixing by imposing the conditions
\be
u_j > 0
\qquad
\forall j.
\label{3.38}\ee
For fixed $\hat p$, the positive vector $u$ remains arbitrary, and thus it can be
uniquely parametrized by introducing a new variable $\hat q \in \bR^n$ via
equation (\ref{3.24}).

The outcome of the above discussion is that the manifold $S_2$ given by
(\ref{3.27}) is a global cross section of the $K$-orbits
in $(\Phi^\ext)^{-1}(0)$, which provides a model of the
reduced phase space (\ref{3.5}).
Indeed, the formula (\ref{3.33}) of $g_-^2$ becomes identical to the formula (\ref{3.23})
of $L_2$ if we take into account the gauge fixing conditions (\ref{3.38}) and
the parametrization (\ref{3.24}).
At the same time, the inversion of the relation (\ref{3.32}) yields  $v(\hat p, \hat q)$ in (\ref{3.25}).

It remains to show that the variables
$(\hat p, \hat q)$ that parametrize $S_2$ are Darboux coordinates on the reduced phase space.
Direct substitution into the symplectic form $\Omega^\ext$ (\ref{3.2}) is now cumbersome, since
we do not have $L_2(\hat p, \hat q)^{\frac{1}{2}}$ explicitly.
We circumvent this problem by proceeding as follows.
We define the smooth functions $F^a$ and $E^a$ on $T^*G \times \cO_\kappa^L$ by
\be
F^a(g, J^R, \xi):= \frac{1}{a} \tr [(J_-^R)^a]
\quad\hbox{and}\quad
 E^a(g,J^R,\xi):= \frac{1}{a} \tr [(g g^\dagger)^a],
\qquad
a=1,\ldots, n.
\label{3.39}\ee
We also define the functions $\cF^a$ and $\cE^a$ on the gauge slice $S_2$ by
\be
\cF^a(\hat p, \hat q):= \frac{1}{a}\tr [\mathbf{\hat p}^a]
\quad\hbox{and}\quad
 \cE^a(\hat p, \hat q):= \frac{1}{a}\tr [L_2(\hat p, \hat q)^a],
\qquad
a=1,\ldots, n.
\label{3.40}\ee
It is clear that $F^a$ and $E^a$ are $K$-invariant functions, and
by means of the  injection map $\iota_2: S_2 \to T^*G \times \cO_\kappa^L$  we have
\be
\iota_2^* F^a = \cF^a,
\qquad
  \iota_2^* E^a = \cE^a \qquad
\forall a=1,\ldots, n.
\label{3.41}\ee
Therefore,  we can determine the induced Poisson brackets of these functions by two methods.
First, denote the Poisson bracket on the extended phase space by $\{.,.\}^\ext$.
The Poisson bracket of $K$-invariant functions is again $K$-invariant
and a straightforward calculation gives (for any $1\leq a,b \leq n$)
\be
\iota_2^* \{ E^a, E^b\}^\ext = \iota_2^* \{ F^a, F^b\}^\ext =0,
\qquad
\iota_2^* \{ E^a, F^b\}^\ext = 2 \,\tr[(\mathbf{\hat p})^{b-1}L_2^a].
\label{3.42}\ee
Second, denote by $\{.,.\}^\red$ the Poisson bracket on $C^\infty(S_2)$ induced by
the reduced symplectic structure. We wish to show that
\be
\{ \hat p^j, \hat p^k\}^\red = \{ \hat q_j, \hat q_k\}^\red =0
\quad\hbox{and}\quad
\{ \hat p^j, \hat q_k\}^\red = \delta^j_k.
\label{3.43}\ee
Now, if we assume that the last relation holds, then it can be
verified by direct
calculation\footnote{Only the relation $\{ \cE^a, \cE^b\}^\red=0$ requires non-trivial effort,
but this was established in \cite{RS}  as well as in
the more recent papers dealing with the  dynamical $r$-matrix structure of the
Ruijsenaars-Schneider models \cite{Avan}.}
(for any $1\leq a,b \leq n$) that
\be
 \{ \cE^a, \cE^b\}^\red = \{ \cF^a, \cF^b\}^\red =0,
\qquad
\{\cE^a, \cF^b\}^\red = 2 \,\tr[(\mathbf{\hat p})^{b-1}L_2^a].
\label{3.44}\ee
From general principles, the restriction of the Poisson bracket of $K$-invariant functions
to a gauge slice always yields the induced Poisson bracket of the restricted functions.
By taking into account (\ref{3.41}),
we conclude from the comparison of the equations (\ref{3.42}) and (\ref{3.44}) that the Poisson bracket
on $C^\infty(S_2)$  that arises from the symplectic reduction coincides with
the Poisson bracket given in coordinates by (\ref{3.43}), at least if we restrict our attention to the
collection of the functions $\cF^a$, $\cE^a$. Thus it remains to prove that the functions  $\cF^a$, $\cE^a$
can serve as local coordinates  around any point  from a dense  open submanifold of $S_2$.
Indeed, the symplectic
form on this dense  submanifold could be then written as $\sum_k d\hat q_k\wedge d\hat p^k$.
The smoothness of the symplectic form  would then allow us to
conclude the same on the whole of $S_2$.

The map from $\C \times \bR^n$ to $\C\times (\bR_+)^n$ given  by
$(\hat p, \hat q)\mapsto (\hat p, u(\hat p, \hat q))$ is clearly a diffeomorphism, and thus
we can use $\hat p^k, u_k$ ($k=1,\ldots, n$) as coordinates
on $S_2$.
When expressed in these new coordinates,
we denote our functions of interest as $\tilde \cF^a$, $\tilde \cE^a$ and $\tilde L_2$:
\be
\tilde \cF^a(\hat p, u)= \cF^a(\hat p, \hat q),
\quad
\tilde \cE^a(\hat p, u)= \cE^a(\hat p, \hat q),
\quad
\tilde L_2(\hat p, u)= L_2(\hat p, \hat q)
\label{tilded-funct}\ee
if $u=u(\hat p, \hat q)$.
To finish the proof, it is sufficient to show that the Jacobian determinant of the map
$\C \times (\bR_+)^n \to \bR^{n}\times \bR^n$ given in coordinates by
 $ \hat p^k, u_k \mapsto \tilde \cF^a(\hat p,u), \tilde \cE^a(\hat p, u)$
is non-vanishing on a dense open subset of $\C \times (\bR_+)^n$.
For this, notice from (\ref{3.23}) and (\ref{3.40}) that  $\tilde \cF^a$ and $\tilde \cE^a$
are rational functions
of the variables $\hat p^k, u_k$, and thus
the Jacobian
\be
\det \frac{\partial (\tilde \cF^a, \tilde \cE^b)}{\partial (\hat p^j, u_k)}
\label{3.45}\ee
is also a  rational  function of the same variables
(that is, the quotient of two polynomials in the $2n$-variables $\hat p^k, u_k$).
This implies that the Jacobian (\ref{3.45})  either vanishes
identically or is non-vanishing on a dense open subset  of $\C\times (\bR_+)^n\simeq S_2$.

Now we show that the Jacobian  (\ref{3.45}) does not vanish identically.
First of all, from the fact that $\tilde \cF^a$ does not depend on $u_k$,
we see easily that
\be
\det \frac{\partial (\tilde \cF^a, \tilde \cE^b)}{\partial (\hat  p^j, u_k)}=
\det\biggl[\frac{\partial  \tilde \cF^a}{\partial \hat p^j}\biggr]
\det\biggl[\frac{\partial  \tilde \cE^b}{\partial u_k}\biggr].
\label{dcp}\ee
The first determinant on the r.h.s.
  is the Vandermonde one:
\be
\det\biggl[\frac{\partial  \tilde \cF^a}{\partial \hat  p^j}\biggr]=\prod_{i<m}(\hat  p^m-\hat p^i),
\label{vdm}\ee
which never vanishes on $S_2$  due to (\ref{3.35}).
The second determinant on the r.h.s of (\ref{dcp})
parametrically depends on $\hat p$. In particular, if we take a real parameter
$s>0$ and consider the curve $\hat p(s)$ defined by
$\hat p^j(s):=e^{(n+1-j)s}$ for  $j=1,\ldots,n$, then we observe that in the limit $s\to\infty$
the matrix $\tilde L_2$ (\ref{tilded-funct}) becomes diagonal,
\be
\lim_{s \to \infty}  \tilde L_2( \hat p(s), u)=\diag(u_1^2,...,u_n^2).
\ee
Hence, in the limit $s\to\infty$, we encounter  again a Vandermonde determinant:
\be
\lim_{s\to\infty}
\det\biggl[\frac{\partial  \tilde \cE^b}{\partial u_k}\biggr]( \hat  p(s), u)
=
2^n \left(\prod_j u_j \right) \prod_{i<m}(u_m^2-u_i^2).
\label{abc}\ee
 Obviously, we can choose $u$ in such a way that the r.h.s. of equation (\ref{abc})
 does not vanish.
 Then the Jacobian determinant (\ref{3.45})  does not vanish at
$(\hat p(s), u)$ for large enough $s$.
 \emph{Q.E.D.}

\medskip
\noindent
\textbf{Remark 2.}  The consequence (\ref{3.31}) of the moment map constraint yields
the commutation relation satisfied by the Lax matrix $L_2$ (\ref{3.23}) and $u(\hat p, \hat q)$ (\ref{3.24}):
\be
[ L_2(\hat p, \hat q), \mathbf{\hat p}] +
2\ri \kappa \left( u(\hat p,\hat q)u(\hat p,\hat q)^\dagger - L_2(\hat p, \hat q) \right) =0,
\ee
which played a crucial role in the analysis in \cite{SR-CMP}.

\section{The dual pair of many-body models}
\setcounter{equation}{0}

We now enumerate important consequences of the results presented in the
preceding sections.

\begin{enumerate}

\item{Since  $S_1$
(\ref{3.15}) and $S_2$ (\ref{3.27})
are two models of the same reduced phase space (\ref{3.5}),
there exists a \emph{natural symplectomorphism} between these two models:
\be
(S_1, \sum_k dp_k \wedge dq^k)\equiv
(T^*G \times \cO_\kappa^L//_0 K, \Omega^\red) \equiv (S_2, \sum_k d\hat q_k \wedge d\hat p^k).
\label{4.1}\ee
By definition, a point of $S_1$ is related to that point of $S_2$ which  represents the
same element of the reduced phase space.}

\item{The pull-backs of the `unreduced Lax matrices' (\ref{3.10+}) to $S_1$ and $S_2$ satisfy, respectively,
\be
\iota_1^* \cL_1^\ext = L_1
\quad\hbox{and}\quad
\iota_2^* \cL_2^\ext =L_2.
\label{4.2}\ee
By the symplectic reduction,
the $K$-invariant Hamiltonians (\ref{3.11+}) descend  to the families
of Hamiltonians defined on $(S_1, \sum_k dp_k \wedge dq^k)$ and on
$(S_2, \sum_k d\hat q_k \wedge d\hat p^k)$, respectively, by
\be
H_j^\red =\frac{1}{j} \tr(L_1^j),
\quad (j=1,\ldots, n)
\quad\hbox{and}\quad
\hat H_k^\red = \frac{1}{2k} \tr(L_2^k),
\quad (k=\pm 1,\ldots, \pm n).
\label{4.3}\ee
The commutativity of $\{ H_j^\red\}$ and $\{ \hat H_k^\red\}$ (where the elements
with fixed sign of $k$ form a complete set) is inherited from the commutativity of the unreduced
Hamiltonians in (\ref{3.11+}). (Note that
$\tr (L_1^j)=\Re \tr( L_1^j)$ since
$L_1$ (\ref{3.14}) is a Hermitian matrix.) }

\item{According to classical results,  $L_1$ (\ref{3.14}) is the Lax matrix of the
hyperbolic Sutherland  model  and $L_2$ (\ref{3.23}) is the Lax matrix of the rational
Ruijsenaars-Schneider model.
The  basic Hamiltonians of these many-body models are indeed reproduced as
\be
 H_{\mathrm{hyp-Suth}}(q,p) \equiv \frac{1}{2}\sum_k p_k^2 + \frac{\kappa^2}{2}\sum_{j\neq k}
\frac{1}{\sinh^2(q^j-q^k)}=
 \frac{1}{2}\tr(L_1(q,p)^2),
 \label{4.4}\ee
 \be
H_\mathrm{rat-RS}(\hat p, \hat q) \equiv  \sum_k \cosh(\hat q_k) \prod_{j\neq k}\left[ 1+
\frac{4\kappa^2}{(\hat p^k - \hat p^j)^2}\right]^\frac{1}{2}
=
\frac{1}{2}\tr(L_2(\hat p, \hat q) +  L_2(\hat p, \hat q)^{-1}).
\label{4.5} \ee
The Lax matrices themselves arose naturally (\ref{4.2}) by means
of the symplectic reduction.}

\item{Consider two points of $S_1$ and $ S_2$ that are related by the
symplectomorphism (\ref{4.1}).
Suppose that these two points are parametrized by $(q,p)\in \C\times \bR^n$ and by
$(\hat p, \hat q)\in \C\times \bR^n$ according to (\ref{3.15}) and (\ref{3.27}), respectively.
The fact that they lie on the same $K$-orbit means, since $g_+$ in $g=g_- g_+$ (\ref{3.17}) is fixed to $\1_n$
in both gauges, that there exists some $\eta\in U(n)$ (actually unique up to the center
of $U(n)$ that acts trivially) for which
\be
(\eta e^{\mathbf{q}} \eta^{-1}, \eta L_1(q,p) \eta^{-1}, -\eta \mu_\kappa \eta^{-1})
= (L_2(\hat p, \hat q)^\frac{1}{2},\mathbf{\hat p}, \xi(\hat p,\hat q)).
\label{4.6} \ee
This shows that the matrix $\mathbf{\hat p}$, which encodes the coordinate-variables
of the rational Ruijsenaars-Schneider model, results by diagonalizing the Sutherland Lax matrix
$L_1(q,p)$.
The same formula shows that $e^{2 \mathbf{q}}$, which encodes the
coordinate-variables of the hyperbolic Sutherland model,  results by diagonalizing
the  Ruijsenaars-Schneider Lax matrix $L_2(\hat p, \hat q)$.
The original direct construction \cite{SR-CMP} of the map between the phase spaces of the two many-body models
relied on diagonalization of the Lax matrices, but in that approach it was quite difficult
to prove the canonicity of the map, which comes for free in the symplectic reduction framework.
}

\item{
It is obvious from the above observations that the two many-body models characterized
by the Hamiltonians (\ref{4.4}) and (\ref{4.5}) are dual to each other in the sense
described in the Introduction.
On the one hand, the Ruijsenaars-Schneider particle-coordinates $\hat p^1,\dots, \hat p^n$
\emph{regarded as functions on $S_1$} define action variables for the hyperbolic Sutherland model.
On the other,
the Sutherland particle coordinates $q^1,\ldots, q^n$ \emph{regarded as functions on $S_2$}
can serve as action variables for the rational Ruijsenaars-Schneider model.
}

\item{The  well-known solution algorithms \cite{OP-Rep,RS} for the commuting
Hamiltonians exhibited in (\ref{4.3})  can be viewed as byproducts  of the
geometric approach.
First of all, it should be noted that all the flows generated by the reduced Hamiltonians
are \emph{complete}, since this is true for the unreduced Hamiltonians whose flows stay in
$(\Phi^\ext)^{-1}(0)$.
By taking an initial value on the gauge slice $S_1$ and projecting the flow (\ref{2.8}) back to
$S_1$ we obtain that the reduced Hamiltonian $H_j^\red$ (\ref{4.3}) generates the following evolution
for the Sutherland coordinate-variables:
\be
e^{2 \mathbf{q}(t)} = \cD[ e^{\mathbf{q}(0)}\exp( 2 t L_1(0)^{j-1}) e^{\mathbf{q}(0)}],
\label{4.7}\ee
where the zero argument refers to the $t=0$ initial value, and $\cD$ denotes the operator
that brings its Hermitian matrix-argument to diagonal form with eigenvalues in non-increasing order.
Similarly, we obtain from (\ref{2.9}) that the reduced Hamiltonian $\hat H_k^\red$ (\ref{4.3}) generates
the following flow for the Ruijsenaars-Schneider coordinate variables:
\be
\mathbf{\hat p}(t)= \cD[\mathbf{\hat p}(0) - t L_2(0)^k].
\label{4.8}\ee
We here used  that $(g^\dagger(0)g(0))^k=
L_2(0)^k$ for any initial value on $S_2$.
}

\end{enumerate}

To summarize, we presented
the duality between the hyperbolic Sutherland model (\ref{4.4})
and the rational Ruijsenaars-Schneider model (\ref{4.5}) in the framework of
 symplectic reduction.
In this way we obtained a Lie theoretic understanding of results
due to Ruijsenaars \cite{SR-CMP}, who originally discovered and investigated the duality by
direct means. Our approach  also simplifies a considerable portion of the original technical arguments.
The general line of reasoning that we followed
may be adapted to explore more complicated cases of
the duality in the future, too.
For example, it will be demonstrated  in \cite{FK1,FK3} that
the reduction approach works in a conceptually very  similar
manner  for the  dualities concerning trigonometric
  Ruijsenaars-Schneider models.

Finally, let us mention that the dual pairs of models studied by
Ruijsenaars at the classical level \cite{SR-CMP,RIMS94,RIMS95}
are associated at the quantum mechanical level with so-called  bispectral problems \cite{DG},
as was first conjectured in \cite{SR-Kup}  and later confirmed in several papers.
Concerning the bispectral property, and in particular
the bispectral interpretation of the duality
between the hyperbolic Sutherland and the rational Ruijsenaars-Schneider models,
the reader may consult \cite{Chal,Haine} and references therein.
We expect that this intriguing phenomenon could be understood also in terms of a
quantum Hamiltonian reduction counterpart of our approach.

\bigskip
\bigskip
\noindent{\bf Acknowledgements.}
L.F. was partially supported
by the Hungarian
Scientific Research Fund (OTKA grant
 T049495) and by the EU network `ENIGMA'
(MRTN-CT-2004-5652).
He wishes to thank to I. Marshall and V. Rubtsov for useful discussions and remarks.

\end{document}